Spin-dependent transport in nanocomposite C:Co films

Shengqiang Zhou[a] , , Markus Berndt[a], Danilo Bürger[a], Viton Heera[a], Kay Potzger[a], Gintautas Abrasonis[a], György Radnóczi[b], György J. Kovács[a], Andreas Kolitsch[a], Manfred Helm[a], Jürgen Fassbender[a], Wolfhard Möller[a] and Heidemarie Schmidt[a]

[a]Institut für Ionenstrahlphysik und Materialforschung, Forschungszentrum Dresden-Rossendorf, Bautzner Landstraße 128, 01328 Dresden, Germany

[b]Research Institute for Technical Physics and Materials Sciences HAS, Konkoly-Thege Mút 29-33, 1121 Budapest, Hungary

## Abstract

The magneto-transport properties of nanocomposite C:Co (15 and 40 at.% Co) thin films are investigated. The films were grown by ion beam co-sputtering on thermally oxidized silicon substrates in the temperature range from 200 to 500 °C. Two major effects are reported: (i) a large anomalous Hall effect amounting to 2 μΩ cm, and (ii) a negative magnetoresistance. Both the field-dependent resistivity and Hall resistivity curves coincide with the rescaled magnetization curves, a finding that is consistent with spin-dependent transport. These findings suggest that C:Co nanocomposites are promising candidates for carbon-based Hall sensors and spintronic devices.

## 1. Introduction

The manipulation of electrical transport in materials through applied magnetic fields is a key feature of spintronic devices. Carbon-based materials have emerged as versatile choices for spintronic applications due to their high electron mobility and low atomic number [1] and [2]. Recently, there has been extensive theoretical and experimental demonstration of spin-dependent transport in graphite and graphene [2], [3] and [4] using ferromagnetic metal contacts. On the other hand, transition metal (TM) nanoparticles encapsulated in a carbon matrix (C:TM) have attracted considerable attention in the past decade due to their unique mechanical, tribological and magnetic properties [5], [6], [7], [8], [9] and [10]. However, there has been little investigation of their magneto-transport properties. We focus on the exploitation of the spin polarization in carbon nanocomposites for use in future spintronic devices. A small negative magnetoresistance (MR) effect has been observed in C:Co [8] and C:Ni [11] and [12] nanocomposites. The scattering by magnetic grains as the MR mechanism has been excluded in Refs. [8] and [11]. A very large negative MR up to 59% at 3 K and 90 kOe has been reported in amorphous Ni-doped $CN_x$ films.

The MR effect is attributed to a spin-related high-order tunneling process between Ni-rich particles. However, no correlation between the magnetization and the MR effect has been provided, which is expected for the tunnel magnetoresistance effect [13]. On the other hand, a large anomalous Hall effect (AHE) has usually been observed in nanogranular magnet–insulator [14] and [15] and magnet–metal [16] and [17] systems. For a ferromagnetic metal, the anomalous Hall resistivity (AHR) is usually of the order of $10^{-3}$ μΩ cm. In this study we report a large saturated AHR of 2 μΩ cm and a negative MR in nanocomposite C:Co films. Both the field-dependent resistivity and the Hall resistivity curves correspond well to the magnetization curve, which suggests a spin-dependent transport in the granular system. Compared with graphene or nanotubes, the nanocomposite C:TM system provides an easier and cheaper test-bed for spin-dependent transport in carbon-based materials. Although the effect is small, it can be enhanced by tuning the size of TM nanoparticles and by creating a regular assembly of matrix and nanoparticles.

## 2. Experimental procedure

Nanocomposite C:Co thin films were grown by ion beam co-sputtering on $SiO_2$ (∼500 nm)/Si substrates. The experimental setup is described elsewhere [18]. Briefly, a 3 cm Kaufman-type ion source is used to produce an $Ar^+$ ion beam with an ion energy of 1 keV and an ion current of 40 mA which is directed towards a 6 in. pyrolytic graphite target (99.99% purity) partially covered with a Co (99.9% purity) stripe. The metal content in the films can be varied by changing the width of the Co stripe. Co stripes 2.8 and 5.5 mm wide result in a Co content of 15 or 40 at.%, respectively. The substrates were located on a heatable substrate holder facing the target at a distance of ∼14 cm, and C:Co films were grown in the substrate temperature range of 200–500 °C. This temperature range has been selected based on observations in the literature. At low temperatures (<300 °C), non-magnetic cobalt carbide, $Co_2C$, forms, while above 300 °C, magnetic metallic cobalt becomes the dominant phase [19] and [20]. Before each deposition, the target was pre-sputtered for 15 min. A shutter, placed in front of the sample, was then removed without interruption of the sputtering process and the depositions were performed for 60 min for magneto-transport measurements and for 20 min for high-resolution transmission electron microscopy (HRTEM) investigation. The resulting film thickness after 60 min deposition is ∼100 nm. The film areal density and composition were obtained by elastic recoil detection analysis (ERDA) (for details see Ref. [20]). The film morphology was determined by HRTEM using a JEOL 3010 300 kV microscope. Magnetic properties were measured with a superconducting quantum interference device (SQUID, Quantum Design MPMS) magnetometer. The samples were measured with the field perpendicular or parallel to the film. The temperature-dependent magnetization measurement has been carried out in the following way: the sample was cooled in zero field from above room temperature to 5 K. Then a 50 Oe field was applied, and the zero-field cooled (ZFC) magnetization curve was measured with increasing temperature from 5 to 350 K, after which the field-cooled (FC) magnetization curve was measured in the same field from 350 to 5 K with decreasing temperature. Magneto-transport properties were measured using Van der Pauw geometry with a magnetic field applied perpendicular or parallel to the film plane. Fields up to 60 kOe were applied over a wide temperature range from 5 to 300 K.

## 3. Results and discussion

### 3.1. Structural properties

Fig. 1 shows the dependence of morphology on Co content and growth temperature. Note that thin films prepared for the TEM investigation were about a third as thick as those used for magnetic measurements (~100 nm thick). One can see from the cross-section images that the film morphology does not depend on thickness if the film thickness amounts to around 30 nm. Fig. 1a and c are bright-field cross-section HRTEM images of C:Co films with 15 and 40 at.% Co prepared at 300 °C. In the TEM images the darker areas are grains containing Co. With increasing Co concentration, the grains grow larger and become elongated in the direction of the film growth. The distance between individual grains also becomes larger. With increasing growth temperature, the morphology and the chemical state of Co in nanocomposite C:Co thin films changes significantly. Below the growth temperature of 300 °C, $Co_2C$ is the major phase, while above 300 °C metallic Co becomes the dominant phase. Both phases are simultaneously present at the growth temperature of 300 °C, which is confirmed by a plan-view HRTEM image shown in Fig. 2a. The measured lattice spacings of 0.25 and 0.21 nm correspond to $Co_2C$ and metallic Co, respectively. At a high (500 °C) growth temperature, the metallic Co becomes the dominant phase as shown in Fig. 2b, where only lattice spacings of 0.19 and 0.21 nm of metallic Co have been observed. Fig. 1d displays the cross-section TEM image of a C:Co film with 40 at.% Co prepared at 500 °C. In this sample, metallic Co is the dominant phase and has a better crystalline quality than the smaller Co grains in Fig. 1b. The lattice planes of Co grains become more clearly visible. On the other hand, the C matrix is graphite-like, and the curved graphene planes, being mostly perpendicular to the substrate, are surrounding the metal-containing nanograins. In the plan-view TEM image, the structure of the carbon matrix is also clearly visible; curved graphene-like planes are encapsulating the $Co_2C$ and Co nanocrystals (Fig. 2). Details concerning the evolution of morphology and structure will be published elsewhere [21].

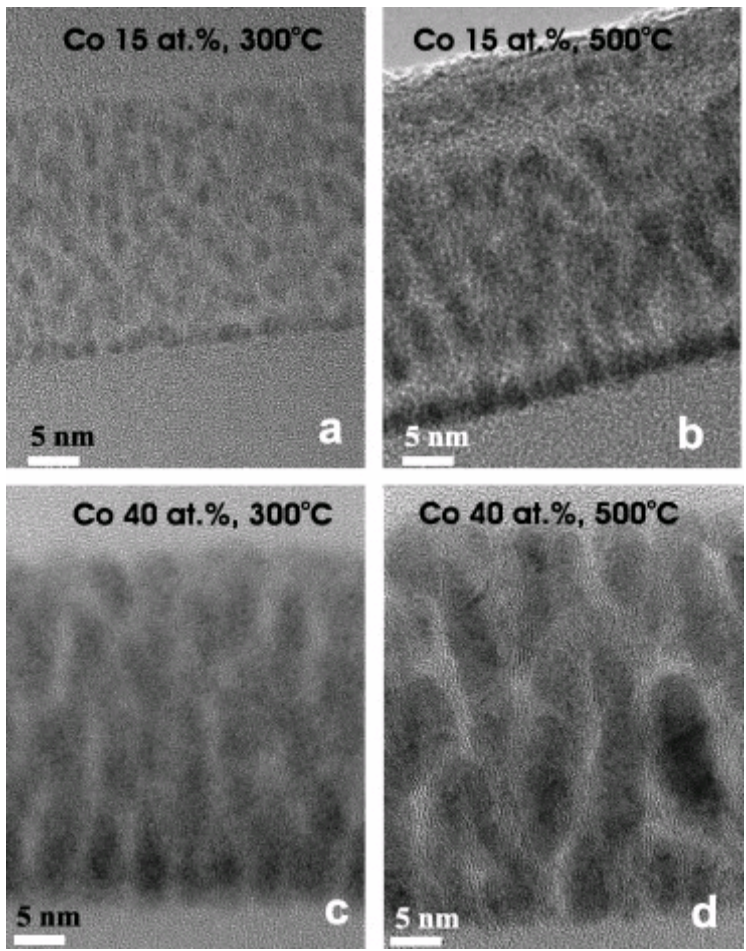

Fig. 1. HRTEM cross-section images of nanocomposite C:Co thin films with different Co concentrations and growth temperatures. The dark areas are Co-rich clusters while white areas are graphite-like carbon. The Co concentration and growth temperature are indicated in the figures. In the sample with 40 at.% Co grown at 500 °C, the crystalline lattice of Co is clearly seen (d).

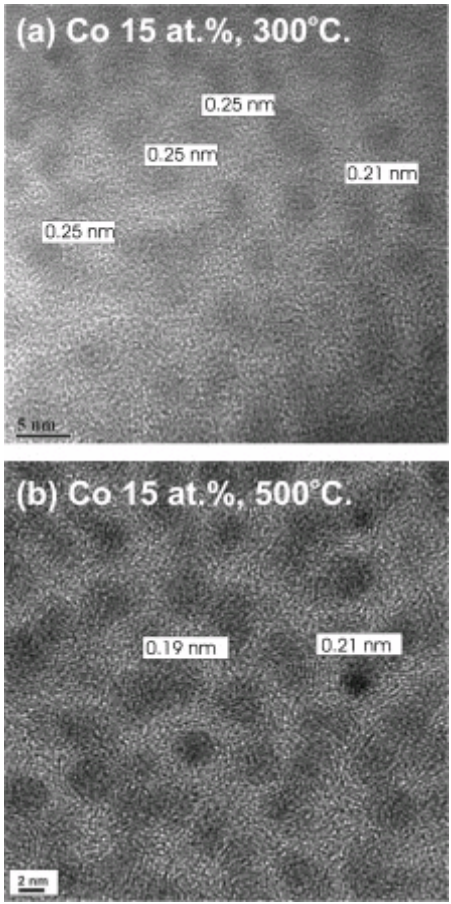

Fig. 2. HRTEM plan-view images of nanocomposite C:Co (15 at.%) thin films grown at (a) 300 °C and (b) 500 °C.

3.2. Magnetic properties

Corresponding to the evolution of the morphology and the phase formation, the magnetic properties of the nanocomposites vary considerably. Fig. 3a shows the ZFC/FC magnetization curves. An irreversible behavior is clearly observed in all ZFC/FC curves. This irreversibility originates from the anisotropy barrier blocking of the magnetization orientation in the nanoparticles cooled under zero field. The magnetization direction of the nanoparticles is frozen as the initial status at high temperature, i.e. randomly oriented. At low temperature (5 K in our case), a small magnetic field of 50 Oe is applied. Some small nanoparticles with small magnetic anisotropy energy flip along the field direction, while the large ones do not. With increasing temperature, the thermal activation energy together with the field flips the larger particles. This process results in an increase in the ZFC curve with temperature. The size distribution of nanoparticles, i.e. the magnetic anisotropy is usually not uniform in the randomly arranged nanoparticle systems. The larger the particles, the higher the anisotropy energy, and a larger thermal activation energy ($k_B T$ with $k_B$ as the Boltzmann constant and $T$ as the temperature) is required for superparamagnetic transition. The temperature at the maximum of the ZFC curve as indicated by arrows in the figure, the blocking temperature $T_B$, increases with growth temperature and Co concentration, i.e. the size of Co grains. The broad peaks in ZFC curves are due to the size distribution of Co grains. Fig. 3b displays the field-dependent magnetization of samples with 40 at.% Co grown at different temperatures. The magnetization increases with growth temperature, which clearly reflects the change in the chemical state of Co. Because $Co_2C$ is non-magnetic, $M_s$ is an indication of the amount of crystalline metallic Co formed in the film [22]. Below the growth temperature of 300 °C, $Co_2C$ is the major phase, while hexagonal close-packed (hcp) Co dominates above 300 °C, which explains the upward jump of the magnetization in the samples grown at 400 °C. By comparing with the metallic face-centered cubic (fcc) or hcp Co with a saturation magnetization of around 1.7 $\mu_B$/Co, we can estimate the fraction of metallic Co in our samples as listed in Table 1, which confirms the results from TEM observations.

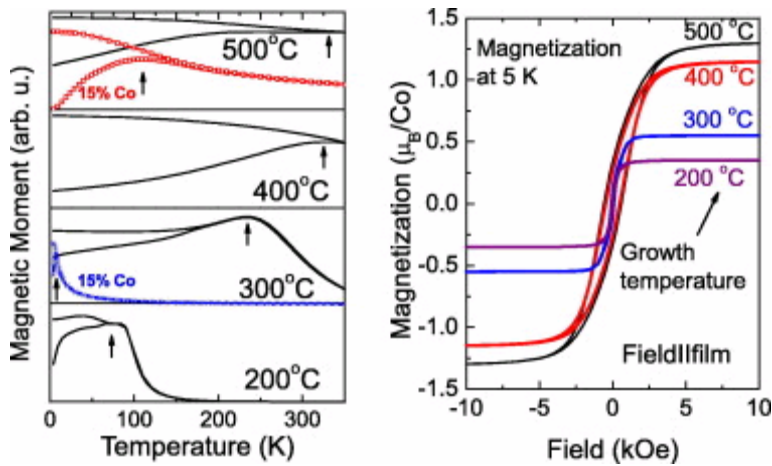

Fig. 3. (a) ZFC/FC (lower branch/upper branch) magnetization curves of nanocomposite C:Co (40 at.%) thin films grown at different temperatures. The blocking temperature (indicated by the arrows), i.e. the size of metallic Co nanoclusters, is increased with increasing growth temperature. Two samples with 15 at.% Co are shown for comparison. (b) Field-dependent magnetization measured at 5 K for nanocomposite C:Co thin films grown at different temperatures. In both cases the applied field is parallel to the film.

Table 1.

Magnetic and magneto-transport properties of nanocomposite C:Co (40 at.%) films grown at different temperatures. All the parameters are measured at 5 K.

| Sub. $T$. (°C) | Magnetization ($\mu_B$/Co) | Resistivity $\rho_{xx}$ (μΩ cm) | Hall resistivity $\rho_{xys}$ (μΩ cm) | MR (%) |
|---|---|---|---|---|
| 200 | 0.35 | 429 | 1.6 | 0.007 |
| 300 | 0.55 | 673 | 2.1 | 0.018 |
| 400 | 1.15 | 1081 | 1.4 | 0.14 |
| 500 | 1.29 | 1590 | 0.7 | 0.18 |

## 3.3. Electrical transport

The temperature-dependent resistivities are displayed in Fig. 4. For all samples the resistivity slightly increases as the temperature decreases as shown in Fig. 4a. The curves exhibit a similar shape, which suggests that the conducting mechanism is similar for all samples. Obviously, the percolation of metallic Co grains can be excluded on the basis of TEM observations and the temperature-dependent resistivity. Hopping conductivity, which is expected for metallic grains embedded inside a dielectric matrix, implies a temperature dependence of $\log(\rho) \sim T^{-0.5}$ [23]. Such a dependence has not been found in the C:Co films of this study. Moreover, the current–voltage curve exhibits a perfect linear behavior, which excludes the tunneling mechanism [11]. On the contrary, the resistivity magnitude and temperature dependence (Fig. 4b) are similar to that of graphitic materials [24] and [25], which shows that the transport of charge carriers takes place through the carbon matrix.

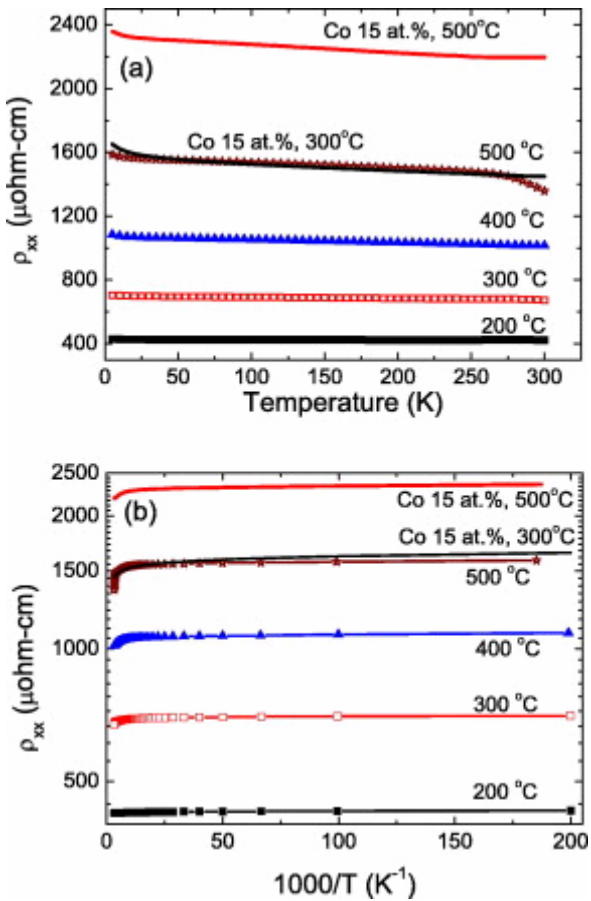

Fig. 4. Temperature dependence of the longitudinal resistivity of nanocomposite C:Co (40 at.%) thin films grown at different temperatures: (a) on temperature and (b) on the reciprocal temperature. Two samples with 15 at.% Co grown at 300 and 500 °C are plotted as solid lines for comparison.

The C:Co films grown at higher temperatures with similar Co content exhibit higher resistivity values. This phenomenon can be explained by the film morphology. At low growth temperatures the nanograins have a globular shape, while the increase in the growth temperature results in the formation of the elongated metal nanoparticles. The graphite-like carbon matrix fills the space between the particles with the graphitic planes following the boundaries of Co nanograins (see Fig. 1c). It is well known that graphite exhibits metallic behavior in the in-plane direction, while in the perpendicular direction it acts as a semiconductor. This is consistent with the semiconducting behavior of C:Co films. In addition, the resistivity increases with growth temperature as a result of the increasing distance between nanograins, which are encapsulated by a graphite-like carbon matrix that is nonconductive in the $c$-direction.

### 3.4. Anomalous hall effect

Fig. 5a shows the field-dependent Hall resistivity ($\rho_{xy}$) measured at 5 K for C:Co composite films with different growth temperatures. During measurements the field is applied perpendicularly to the film. The Hall resistivity

(1)
$$\rho_{xy} = R_0 B + R_s \mu_0 M$$

is known to be the sum of the ordinary and anomalous Hall terms, where $B$ is the magnetic induction, $\mu_0$ is the magnetic permeability, $M$ is the magnetization, $R_0$ is the ordinary Hall coefficient, and $R_s$ is the anomalous Hall coefficient. The ordinary and anomalous Hall terms are linear in $B$ and $M$, respectively. Fig. 5a clearly shows the nonlinear dependence of $\rho_{xy}$ on the field, i.e. the AHE dominates the $\rho_{xy}$ curves. At the low-field part the ordinary Hall effect can be ignored and the Hall resistivity at the saturation field can be considered as the saturation AHR $\rho_{xys}$. In the presence of a large AHE, a very large field is required in order to determine the carrier concentration and the mobility, which is beyond the capability of our setup [26]. It can be seen that $\rho_{xys}$ first increases concomitantly with the film growth temperatures, and reaches a maximum at ~300 °C before decreasing. In bulk ferromagnets $\rho_{xys}$ linearly depends on the sample magnetization. However, in nanocomposite systems one has to consider the enhancement of $\rho_{xys}$ by surface scattering [27]. By comparing the $\rho_{xys}$ observed in granular $SiO_2$:Ni [14] and in $SiO_2$:Co [28], one can also conclude that smaller nanoparticles make a larger contribution to the scattering of conduction electrons than larger ones. In the sample grown at 300 °C, there is a sufficiently large amount of small magnetic Co grains to yield large AHE values. The increase in the growth temperature results in the coarsening of the Co grains, which for constant Co content decreases the Co grain concentration. This consequently reduces the surface scattering effect. Both these effects—an increase in the average grain size and a decrease in their concentration—reduce $\rho_{xys}$. Fig. 5b shows the Hall curves measured at different temperatures for the sample grown at 300 °C. Its magnetization curve measured at 5 K with the field perpendicular to the film is shown for comparison. Obviously, the magnetization curve coincides after scaling with the Hall resistance curve. For superparamagnetic Co nanograins, the magnetization and coercivity decrease with increasing temperature. The hysteresis loops of AHE exhibit the same trend.

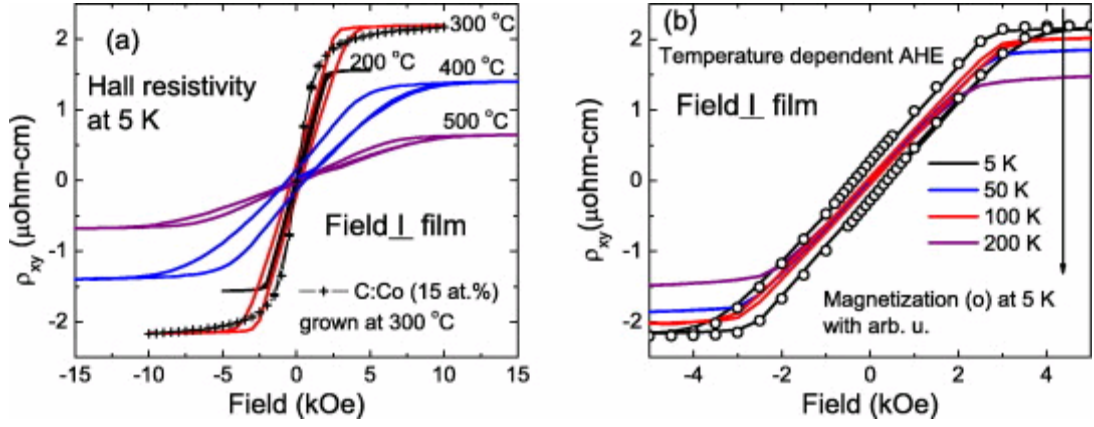

Fig. 5. Field-dependent AHE (a) measured at 5 K for nanocomposite C:Co (40 at.%) thin films grown at different temperatures. The sample C:Co (15 at.%) grown at 300 °C is shown for comparison. (b) AHE measured at different temperatures for the sample grown at 300 °C with 40 at.% Co. The scaled magnetization curve at 5 K (circles) perfectly overlaps with the curve of $\rho_{xy}$. The applied magnetic field is perpendicular to the film.

AHE is believed to be caused by spin–orbit interaction between conduction electrons and lattice disorder such as impurities, phonons, etc. Spin–orbit interaction results in two distinctive mechanisms, i.e. skew scattering and side jump, that affect the Hall effect. Skew scattering causes the electron trajectory to deflect from its original path, while side jump is a pure quantum mechanical effect, in which the electron trajectory is displaced transversely by a small distance of around 0.1–1 Å [16]. To elucidate the scattering mechanism, the correlation between $\rho_{xys}$ and $\rho_{xx}$ has to determined. For bulk ferromagnets $\rho_{xys} \propto \rho_{xx}^n$ and the exponent amounts to $n = 1$ for skew scattering and $n = 2$ for side jump. However, as suggested in previous investigations [29] and [30], the scaling law of the AHE and the resistivity do not have to hold for any composite system because in such systems the scaling parameters have their origin in different scattering centers. It should be noted that the resistivity in the case of C:Co nanocomposites is mostly due to the C matrix. Fig. 6a shows an example of the relation between $\rho_{xys}$ and $\rho_{xx}$ for the sample grown at 400 °C. The exponent $n$ is as large as 2.7. Moreover, as can be seen from Fig. 6b, $n$ increases when the film growth temperature decreases and reaches 7.2 for the samples grown at 200 °C. Such large exponents have been observed in the Co–Ag system ($n = 3.7$) [16] and the Ni–TiN system ($n = 4.7$) [30]. On the other hand, it is not clear what other mechanism could be responsible for the AHE in granular systems. Vedyaev et al. calculated the scaling exponent to be 3.8 for the skew scattering by considering the scattering not only by one grain but by several grains [31]. Their calculations also showed, in agreement with our observation, that scattering of conduction electrons by the grain–matrix interfaces has a substantial effect on the magnitude of the anomalous Hall effect. As can be seen from Fig. 6b, the scaling exponent increases for both Co contents when the film growth temperature decreases, i.e. when the Co grain size decreases. Note that the magnitude of the AHE in the sample grown at 200 °C decreases due to the small amount of ferromagnetic metallic Co (see Table 1).

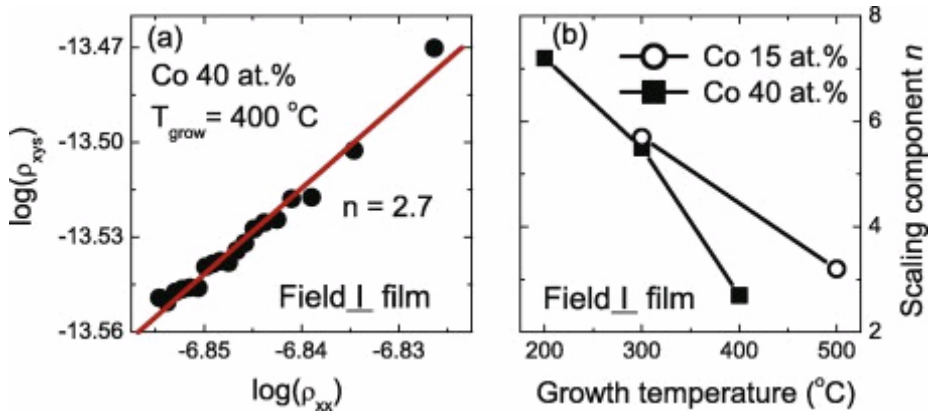

Fig. 6. (a) The correlation between $\rho_{xys}$ and $\rho_{xx}$ for the sample grown at 400 °C in the temperature range from 5 to 100 K. $n$ is the exponent and increases with decreasing growth temperature. (b) The scaling factor $n$ as a function of sample parameters, e.g. the growth temperature and the Co concentration.

### 3.5. Magnetoresistance

Fig. 7a shows the MR curves for the C:Co films grown at different temperatures. It can been seen that MR is negative in the low magnetic field part for all the samples. The magnitude increases with the film growth temperatures, mirroring the trend of the magnetization (Fig. 3b). The magnitude is similar to that reported in the literature for nanocomposite C:Ni [11] and C:Co [8] films. The literature attributes this effect to anisotropic magnetoresistance [11] or to some kind of unknown effect [8]. Despite the low MR value, the butterfly shape of the MR curves of the C:Co films of this study grown at higher temperatures suggests that this effect has to be attributed to the magnetic scattering by Co grains, the so-called giant magnetoresistance (GMR) effect, i.e. the spin-dependent scattering of conduction electrons on Co nanograins. This conclusion is further supported by the comparison of the field-dependent MR and the square of the normalized magnetization $-(M/M_s)^2$ as shown in Fig. 7b and d with the applied field perpendicular and parallel to the film, respectively. First, a negative MR effect is observed with a comparable amplitude for different field orientations, which excludes the anisotropic MR [11]. Second, the square of the normalized magnetization, $-(M/M_s)^2$, correlates reasonably with the field-dependent MR curve. This shows that scattering decreases at the saturation magnetization when the magnetic moment directions in different Co grains are correlated at maximum level. Moreover, a clear hysteresis induced by the coercive force of Co during the upward and the downward sweeps of the applied field is also observed in MR curves as shown in Fig. 7b and d. The coercivities for both magnetic field orientations are similar to those in the MR curves. Such a small GMR effect has also been observed in SiO$_2$:FeSn [32] and SiO$_2$:NiFe [33] composites and NiFe/Ag multilayer thin films [34]. The magnitude of the MR in the sample grown at 500 °C is weakly dependent on temperature (not shown), as is the magnetization. This observation also excludes magnetic tunneling as the MR mechanism [35]. Note that magnetization and MR curves coincide better at the low-field part if the magnetic field is applied along the easy axis (Fig. 7d). When the field is applied along the hard axis (perpendicular to the film) the MR and $-(M/M_s)^2$ do not coincide when the magnetic field is switched (the inner part in Fig. 7b). The MR effect is switched at a larger field than the magnetization, a finding that we do not yet completely understand. It could be due to the abnormal magnetization anisotropy as shown in Fig. 7c. The direction perpendicular to the film is the hard axis, though with a larger coercivity field. In larger fields the slow saturation of the MR curve (Fig. 7d—also in 7b but at much larger field, not shown here) may be due to the existence of the paramagnetic component, which arises from the very small Co grains.

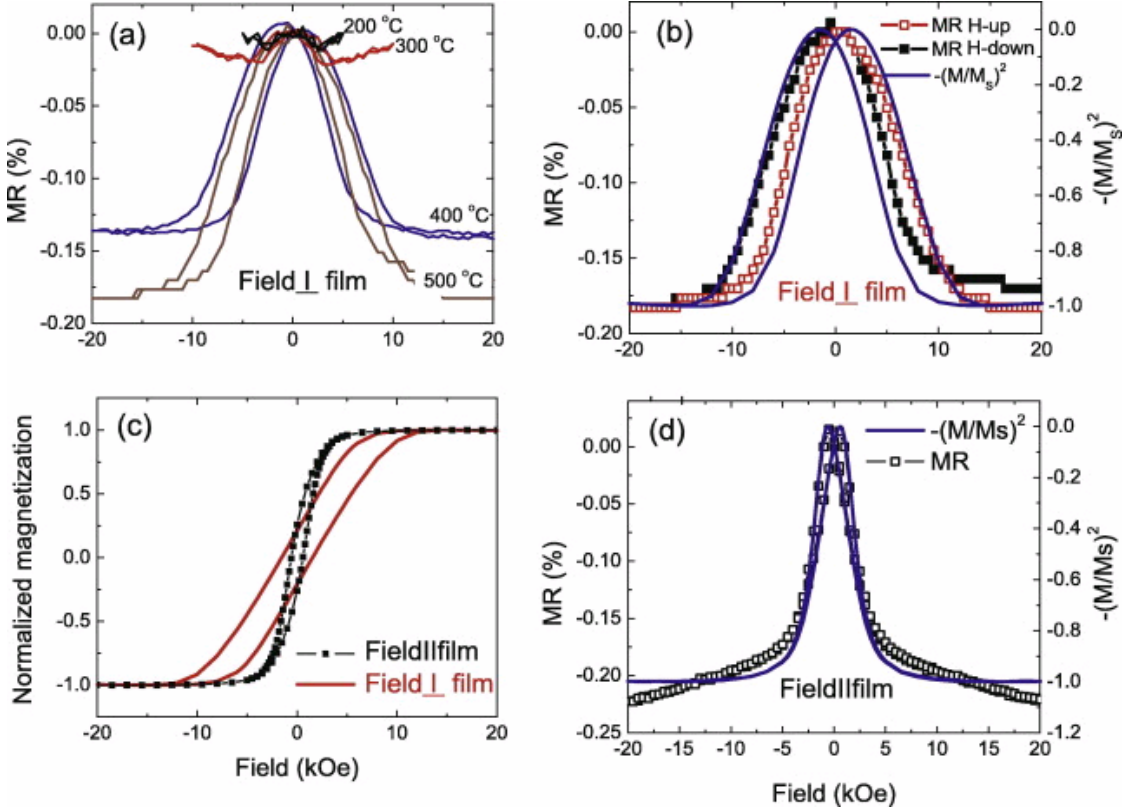

Fig. 7. Field dependencies of resistivity ($\rho_{xx}$), measured at 5 K (a) for samples grown at different temperatures with 40 at.% Co, (b) for the sample grown at 500 °C with the field perpendicular to the film and (d) for the sample grown at 500 °C with the field parallel to the film. In (c) the magnetization measured with the field perpendicular and parallel to the film is shown for comparison. In (b) and (d) the plot of $-(M/M_s)^2$ as a function of the field is shown to compare with the MR curves. The field-dependent resistivity curves correspond reasonably well to the magnetization curve, which suggests a spin-dependent transport in the granular system.

In Refs. [16] and [33] the authors reported the correlation between the GMR and the AHE magnitude. Both values decrease with increasing magnetic grain size. In our case, such a correlation cannot be established (see Table 1), indicating that the models for metal–metal composites and $SiO_2$:TM are not valid for C:TM systems. The elucidation of the spin-scattering mechanisms requires further investigations and will be addressed in future studies.

## 4. Conclusions

In summary, magneto-transport properties of nanocomposite C:Co (15 and 40 at.%) thin films grown by ion beam co-sputtering in the temperature range 200–500 °C have been investigated. Large saturated AHE and a negative MR have been observed, which is consisted with the spin-dependent scattering of conduction electrons by magnetic Co nanograins. The saturation AHE is comparable to that of Co-doped $TiO_2$ [36] and [37] and of Si-based magnetic semiconductors [38]. Graphitic materials have a large electron mobility and a low atomic number which minimize the hyperfine interaction of the electron spins with the nuclei. As these properties strongly depend on the film morphology, they can be optimized by varying the film composition, metal type, substrate temperature, as well as depositing flux energetics for the application of such films as anomalous Hall sensors, memory and logic devices [29], [38] and [39]. In particular, anomalous Hall sensors made from nanocomposite C:Co films may be used for different frequency ranges by tuning the film resistivity.

## Acknowledgements

S.Z. and H.S. acknowledge financial funding from the Bundesministerium für Bildung und Forschung (FKZ03N8708). This work has been carried out as a part of the integrated EU project "Fullerene-based Opportunities for Robust Engineering: Making Optimised Surfaces for Tribology" and is supported by the EU Contract No. NMP3-CT-2005-515840.